\documentclass[sigconf,nonacm]{acmart}
\setcopyright{none}

\AtBeginDocument{%
  }

\usepackage{svg}
\usepackage{textcomp}
\usepackage{xcolor}
\usepackage{fancyhdr}
\usepackage{algorithm}
\usepackage{algpseudocode} 
\usepackage{algorithmicx}
\usepackage{tabularx}
\usepackage{adjustbox}
\usepackage{subcaption}
\usepackage{makecell}
\usepackage{multirow}
\usepackage{bm}
\usepackage{dblfloatfix}
\algnewcommand{\LineComment}[1]{\State \(\triangleright\) #1}
\usepackage{pifont}
\usepackage{enumitem}
\usepackage{soul}
\usepackage{color,soul}
\usepackage[normalem]{ulem}

\begin{document}
\title{Scalable Processing-Near-Memory for 1M-Token LLM Inference: CXL-Enabled KV-Cache Management Beyond GPU Limits}

\author{Dowon Kim}\authornote{These authors contributed equally to this work.}
\affiliation{\institution{Hanyang University}\country{Republic of Korea}}\email{kdw4537@hanyang.ac.kr}
\author{MinJae Lee}\authornotemark[1]
\affiliation{\institution{Hanyang University}\country{Republic of Korea}}\email{lmj4666@hanyang.ac.kr}
\author{Janghyeon Kim}
\affiliation{\institution{Hanyang University}\country{Republic of Korea}}\email{kkt20@hanyang.ac.kr}
\author{HyuckSung Kwon}
\affiliation{\institution{Hanyang University}\country{Republic of Korea}}\email{momarom@hanyang.ac.kr}
\author{Hyeonggyu Jeong}
\affiliation{\institution{Hanyang University}\country{Republic of Korea}}\email{hgjeong753@hanyang.ac.kr}
\author{Sang-Soo Park}
\affiliation{\institution{Samsung Electronics}\country{Republic of Korea}}\email{ss23.park@samsung.com}
\author{Minyong Yoon}
\affiliation{\institution{Samsung Electronics}\country{Republic of Korea}}\email{casper.yoon@samsung.com}
\author{Si-Dong Roh}
\affiliation{\institution{Samsung Electronics}\country{Republic of Korea}}\email{sidong.roh@samsung.com}
\author{Yongsuk Kwon}
\affiliation{\institution{Samsung Electronics}\country{Republic of Korea}}\email{yssh.kwon@samsung.com}
\author{Jinin So}
\affiliation{\institution{Samsung Electronics}\country{Republic of Korea}}\email{jinin.so@samsung.com}
\author{Jungwook Choi}\authornote{Corresponding author.}
\affiliation{\institution{Hanyang University}\country{Republic of Korea}}\email{choij@hanyang.ac.kr}

\begin{abstract}
The expansion of context windows in large language models (LLMs) to multi-million tokens introduces severe memory and compute bottlenecks, particularly in managing the growing Key-Value (KV) cache. While Compute Express Link (CXL) enables non-eviction frameworks that offload the full KV-cache to scalable external memory, these frameworks still suffer from costly data transfers when recalling non-resident KV tokens to limited GPU memory as context lengths increase. This work proposes scalable Processing-Near-Memory (PNM) for 1M-Token LLM Inference, a CXL-enabled KV-cache management system that coordinates memory and computation beyond GPU limits. Our design offloads token page selection to a PNM accelerator within CXL memory, eliminating costly recalls and enabling larger GPU batch sizes. We further introduce a hybrid parallelization strategy and a steady-token selection mechanism to enhance compute efficiency and scalability. Implemented atop a state-of-the-art CXL-PNM system, our solution delivers consistent performance gains for LLMs with up to 405B parameters and 1M-token contexts. Our PNM-only offloading scheme (PNM-KV) and GPU–PNM hybrid with steady-token execution (PnG-KV) achieve up to 21.9$\times$ throughput improvement, up to 60$\times$ lower energy per token, and up to 7.3$\times$ better total cost efficiency than the baseline,  demonstrating that CXL-enabled multi-PNM architectures can serve as a scalable backbone for future long-context LLM inference.
\end{abstract}


\maketitle

\section{Introduction}
\label{sec:introduction}

The context window of large language models (LLMs) has expanded rapidly, reaching multi-million tokens in state-of-the-art models such as Llama 4~\cite{openai2024gpt4technicalreport, team2023gemini, Touvron2023Llama2O, anthropic2024sonnet, jiang2023mistral7b,llama4_meta}. This growth enables advanced capabilities such as long document summarization~\cite{zhong-etal-2021-qmsum}, multi-step reasoning~\cite{wei2022chain}, and repository-level code analysis~\cite{liu2023repobench}. However, long-context inference introduces significant system-level challenges. The Key-Value (KV) cache footprint scales linearly with context length and batch size, dominating memory usage by requiring substantial capacity for storing extended sequences and high bandwidth to sustain attention’s matrix-vector multiplications~\cite{gu2025pim}. While GPU architectures have greatly improved compute throughput, their memory capacity and bandwidth have stagnated, with high-bandwidth memory constrained in both size and cost~\cite{ai_memory_wall}. Consequently, scaling conventional GPU systems to meet long-context memory demands is inefficient and unsustainable. This motivates the need for a supplementary system architecture that augments GPU infrastructures by providing scalable memory capacity and high-bandwidth KV-cache access beyond the limitations of traditional GPU memory hierarchies.

Efficient KV-cache management is a central challenge for scaling LLM inference to long contexts. Recent approaches fall into two categories: eviction-based methods, which permanently remove less relevant tokens~\cite{xiao2023efficient,han2023lm,zhang2023h2o,liu2023scissorhands,chen2024nacl,adnan2024keyformer,kim2024infinipot}, and non-eviction methods, which dynamically select relevant tokens without discarding any~\cite{chen2024arkvale,xiao2024infllm,tang2024quest,zhang2024pqcache,hooper2024squeezed,liu2024retrievalattention,fountas2024human,liu2024clusterkv,ribar2023sparq}. Although eviction-based schemes improve memory efficiency, their irreversible nature risks degrading accuracy by discarding useful context. In contrast, non-eviction methods retain the full token history in secondary memory, retrieving a context-dependent subset at inference time to achieve greater adaptability and typically higher accuracy in long-context tasks. System-level implementations~\cite{zhao2024alisa,lee2024infinigen,jiang2024neo,kim2025oaken} offload the KV-cache to CPU memory, but remain constrained by host-side bandwidth.

Compute Express Link (CXL) is emerging as a critical technology for scalable memory systems, offering coherent, low-latency access to external memory beyond the physical limits of GPU-attached DRAM~\cite{sharma2019compute,sharma2022compute}. Recent research demonstrates CXL’s effectiveness in memory pooling, tiered memory management, and production deployments~\cite{li2023pond,gouk2023memory,sun2023demystifying,mao2024cxl,berger2025octopus}. For long-context LLM inference, where KV-cache usage scales linearly with context length and batch size, CXL provides a practical means to extend memory capacity without sacrificing performance. Tang et al.~\cite{tangexploring} show that offloading the KV-cache to CXL memory can reduce GPU memory usage by up to 87\% while meeting latency requirements, and Gouk et al.~\cite{gouk2024breaking} demonstrate sub-100 ns access latency with a CXL memory controller optimized for AI workloads. Unlike CPU-attached memory, which suffers from limited bandwidth and incoherent access, CXL supports low-overhead, byte-addressable communication between GPU and memory expanders, making it ideal for managing large KV-caches in non-eviction LLM frameworks and enabling scalable multi-million-token inference.

These frameworks commonly partition the KV-cache into pages or clusters, each summarized/compressed by compact metadata stored in GPU memory~\cite{xiao2024infllm,tang2024quest,hooper2024squeezed,chen2024arkvale,liu2024clusterkv}. At runtime, query-to-summary/compression similarity matching identifies relevant pages for attention computation. When selected pages are missing from GPU memory, they are \textit{recalled} from offloaded storage—often using CXL memory, which offers sufficient capacity to hold the full KV-cache without eviction. This offloading capability has made non-eviction strategies fundamental for scalable, long-context inference, enabling full-context coverage for multi-million-token inputs. However, as context lengths grow into the hundreds of thousands to millions of tokens, the efficiency of these frameworks declines. First, the number of recall operations increases proportionally with context length (Figure~\ref{fig:challenges}(a)), resulting in significant offloading and retrieval overhead, especially over bandwidth-constrained CXL links. Second, because attention must be computed within the GPU where recalled KV-cache pages reside, GPU memory pressure limits batch size, leading to underutilized compute resources, particularly for fully connected layers that benefit from large batch execution (Figure~\ref{fig:challenges}(b)). These challenges reveal fundamental bottlenecks in current dynamic selection frameworks when scaling to extreme context lengths, highlighting the need for more integrated memory and computation strategies.

To address the inefficiencies of dynamic KV-cache selection under extreme context lengths, we propose a CXL-enabled multi-PNM KV-cache management system that leverages CXL’s composability to enable Processing-Near-Memory (PNM) for scalable and efficient attention computation. First, we implement a PNM accelerator that performs token page selection directly within CXL memory modules, eliminating costly GPU recall. By storing the full KV-cache in CXL memory and offloading page selection to PNM, we relieve GPU memory pressure and support larger batch sizes for fully connected (FC) layers, improving overall inference throughput. Second, we introduce a parallelization strategy optimized for multi-PNM deployment: data parallelism (DP) is applied to PNM-based attention computation to avoid expensive head-wise reductions required in tensor parallelism (TP), while FC computation remains TP-parallelized on the GPU. Hybrid partitioning is enabled by cost-free mapping between TP and DP domains during GPU–PNM coordination, ensuring constant communication overhead regardless of context length. Third, to better utilize idle GPU time during attention processing, we propose a GPU–PNM hybrid parallelization strategy that balances memory capacity and compute efficiency. We introduce a \textit{steady-token selection} algorithm to identify tokens with persistent relevance over time, minimizing data movement from CXL memory to the GPU. By retaining a whole batch of steady tokens on the GPU, we maintain large batch sizes and high FC computation throughput. While prior work has explored programmable logic in CXL-attached devices~\cite{kwon2023failure,tangexploring,gouk2024breaking,park2024cxlpim,ham2024low,ji2024demystifying,gu2025pim}, our approach uniquely targets the deterministic and disaggregated nature of KV-cache access to demonstrate a scalable, parallelizable, and GPU-supplementing PNM architecture. This underscores CXL’s emerging role not only as a memory expander but as a computational backbone for future AI systems.

We implement the proposed CXL-enabled multi-PNM attention accelerator by extending a state-of-the-art CXL-PNM architecture for LLM inference~\cite{park2024cxlpim}, incorporating custom silicon support for dynamic and steady-token selection mechanisms. We also design our CXL-PNM architecture to support two operating models. The first, called PNM-KV, offloads KV-cache management and full attention calculations to CXL-PNM. The second, called PnG-KV, adopts a hybrid execution model that leverages idle GPU resources during attention computation for further optimization. We evaluate our system at both server and rack-scale levels: integrating the PNM accelerator into a GPU server for hybrid execution and deploying standalone PNM nodes within a rack-scale setup. Experiments on LLMs with 7B, 70B, and 405B parameters—at context lengths up to 1 million tokens—demonstrate consistent and scalable performance gains across deployment scales, highlighting the potential of CXL-enabled multi-PNM architectures for long-context LLM inference. Specifically, our PNM-KV and PnG-KV achieve up to 21.9$\times$ throughput improvement, up to 60$\times$ lower energy per token, and up to 7.3$\times$ higher total cost efficiency compared to the baseline, establishing a new scalability and efficiency standard for multi-million-token LLM inference.

\section{Background}
\label{sec:background}

\subsection{Efficient KV-Cache Management}
\label{subsec:background_lllm}

\textbf{Long-Context LLM Inference.}
Long-context LLMs maintain coherence across tens of thousands of tokens, enhancing contextual relevance in tasks such as long document summarization~\cite{zhong-etal-2021-qmsum}, which produces cohesive summaries from dispersed content, and repository-level code analysis~\cite{liu2023repobench}, which enables programming assistants to understand entire codebases. Chain-of-thought (CoT) reasoning further enhances answer quality by supporting multi-step reasoning~\cite{wei2022chain, openai2024o1}. These capabilities depend on inference-time processing of extended contexts, with modern LLMs supporting sequence lengths from 16K to 200K~\cite{bai2023qwen, mpt7b, anthropic2024sonnet, openai2024o1}, and some reaching up to 1M tokens~\cite{team2024gemini, yang2024qwen2}. Table~\ref{tab:LLMs} summarizes representative long-context LLMs, detailing the number of layers ($n_l$), attention heads ($n_h$), per-head feature dimension ($d_h$), hidden dimension ($d_{in}$), intermediate dimension ($d_{out}$), and architectural variants such as group query attention (GQA), where queries share grouped keys and values~\cite{ainslie2023gqa}. These models are trained on long-context datasets to handle tasks requiring extended input processing.

\begin{table}[t]
    \centering
    \begin{adjustbox}{width=\linewidth}
    \begin{tabular}{c|c|c|c|c|c|c}
        \hline
        \makecell{Model}    & $n_l$ & $n_h$ & $d_h$ & $d_{in/out}$  & GQA       & CW    \\ \hline \hline
        Llama3.1-8B         & 32    & 32    & 128   & 4K/14K        & ($g=4$)   & 128K  \\ \hline
        Llama3.1-70B        & 80    & 64    & 128   & 8K/28K        & ($g=8$)   & 128K  \\ \hline
        Llama3.1-405B       & 126   & 128   & 128   & 16K/53K       & ($g=16$)  & 128K  \\ \hline
    \end{tabular}
    \end{adjustbox}
    \vspace{5pt}
    \caption{LLM specification with a group size ($g$) for GQA and the context window (CW).}
    \vspace{-10pt}
    \label{tab:LLMs}
\end{table}

\begin{figure}[h]
    \centering
    \includegraphics[width=1\linewidth]{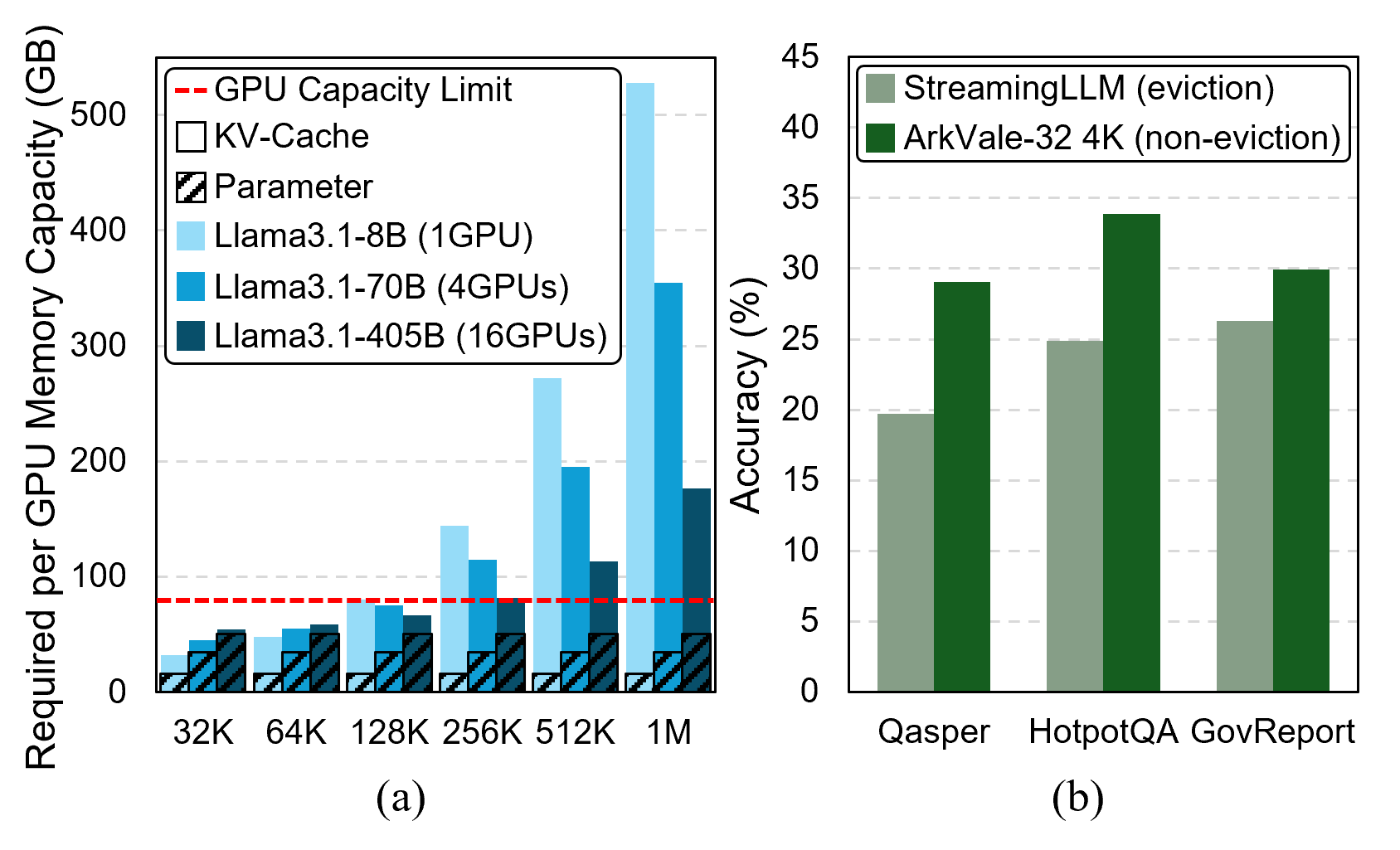}
    \vspace{-23pt}
    \caption{Characteristics of long-context LLM decoding. (a) Per-GPU memory demand increases beyond the GPU capacity limit as context length increases. (b) Accuracy comparison of KV-cache management algorithms: dynamic selection with eviction (StreamingLLM~\cite{xiao2023efficient}) and without eviction (ArkVale~\cite{chen2024arkvale}), evaluated on LongBench~\cite{bai2024longbench} tasks.}
    \vspace{5pt}
    \label{fig:capa}
\end{figure}

\textbf{Capacity Demand for KV-Cache.}
During long-context LLM inference, the KV-cache becomes a major memory overhead. In the Transformer decoder architecture~\cite{vaswani2017attention}, each decoder block generates per-token Key and Value vectors of dimension $d_h$, which are accumulated into the KV-cache with size $\mathbb{R}^{T \times d_h}$ per attention head generated in the prefill phase with context length T. Given $n_l$ layers, $n_h$ heads per layer, and a batch size $B$, the total KV-cache size scales proportionally to $O(B \cdot T \cdot n_l \cdot n_h \cdot d_h)$. As the context length $T$ increases, the KV-cache memory demand grows linearly, significantly outpacing the relatively constant memory footprint of model weights. As shown in Figure~\ref{fig:capa}(a), the KV-cache quickly dominates total memory usage, often exceeding the size of the model weights. For large models, we can shard the model across multiple devices and execute attention and feed-forward layers in parallel using tensor parallelism (TP). The KV cache is partitioned by attention head in line with TP, so each GPU keeps the KV for the attention heads assigned to it. However, even if the model parameters of Llama3.1-405B are sharded across 16$\times$A100-80GB GPUs, the remaining GPU memory is still insufficient to hold a 1M-token KV cache. This rapid memory expansion underscores the need for efficient KV-cache management techniques, which aim to minimize the active memory footprint while maintaining model accuracy for scalable, long-context LLM inference.

\textbf{Efficient KV-Cache Management. }
Efficient KV-cache management is critical for scaling LLM inference to long-context scenarios. Recent work exploits the sparsity of attention distributions, observing that a small subset of tokens typically dominates attention computation at each step~\cite{chen2024arkvale,xiao2024infllm,tang2024quest,hooper2024squeezed}. Dynamic KV-cache selection methods leverage this property by retrieving only the Top-$K$ most relevant tokens, reducing memory access and computational cost while preserving inference quality. Approaches that retain the full KV-cache with dynamic selection (a.k.a. non-eviction methods) have shown favorable accuracy compared to eviction-based strategies~\cite{xiao2023efficient,han2023lm,zhang2023h2o,liu2023scissorhands}, which irreversibly discard potentially useful context (e.g., Figure~\ref{fig:capa}(b)). To enable efficient retrieval, tokens are organized into coarse-grained pages or clusters, summarized via pooled key vectors or centroids~\cite{liu2024clusterkv,hooper2024squeezed}, allowing lightweight query-to-summary matching during inference. While non-eviction dynamic selection mitigates accuracy degradation and improves adaptability, it imposes substantial capacity demands to retain the full KV-cache. This motivates the use of CXL-enabled memory expansion, which provides scalable capacity and near-memory access to support efficient and accurate long-context LLM inference.

\begin{figure}[h]
    \centering
    \includegraphics[width=1\linewidth]{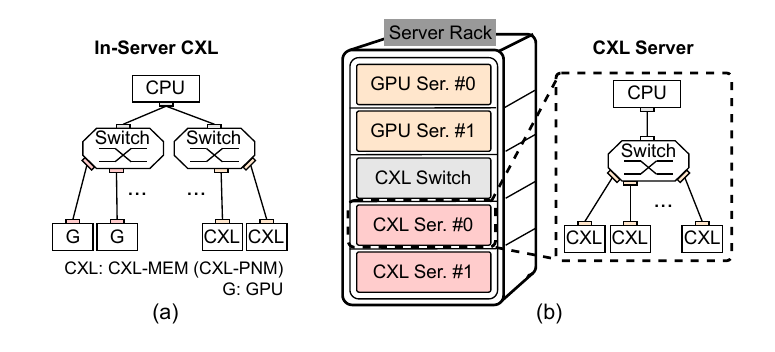}
    \vspace{-23pt}
    \caption{CXL-Enabled Memory Expansion System. (a) A GPU system attached with CXL-Mem/PNM. (b) A rack-scale multi-node GPU system with CXL-enabled memory/PNM nodes.}
    \label{fig:CXL-system2}
\end{figure}

\subsection{CXL-Enabled Memory Expansion}
\label{subsec:background_cxl}

Compute Express Link (CXL) is emerging as a key enabler for disaggregated and tiered memory systems, offering coherent, low-latency access to external memory devices. A significant body of research has investigated the practical realization, performance characteristics, and system-level implications of CXL-based architectures~\cite{li2023pond,gouk2023memory,yang2022performance,zeng2025performance,sun2023demystifying,mao2024cxl,wang2024exploring,yang2025architectural,jang2024bridging,berger2025octopus}. The CXL-enabled memory expanders integrate high-capacity, high-bandwidth DRAM with CXL interfaces to address the memory capacity demand. Specifically, we consider the CXL-enabled memory expansion for LLM inference in two system scales: a GPU system attached with CXL-Mem/PNM and a rack-scale multi-node GPU system with CXL-enabled memory/PNM nodes (Figure~\ref{fig:CXL-system2}). Each CXL device is equipped with LPDDR5X-based CXL memory modules, offering a 512 GB capacity and 1.1 TB/s bandwidth per module, interconnected via a CXL 3.0 switch that supports PCIe 6.0 physical layers and device links, each configured with x8 lanes per device. The host connects to the switch with x16 lanes. The primary communication type is CXL.mem transactions for CPU-to-memory load/store operations (we do not consider CXL's peer-to-peer and collective communication primitives for simple system configuration). These systems significantly expand memory capacity while maintaining a scalable interconnection architecture suitable for large-scale LLM Inference, thereby overcoming the memory limitations of GPU-only systems.

\subsection{Challenges}
\label{subsec:background_challenges}

\begin{figure}[t]
    \centering
    \includegraphics[width=1\linewidth]{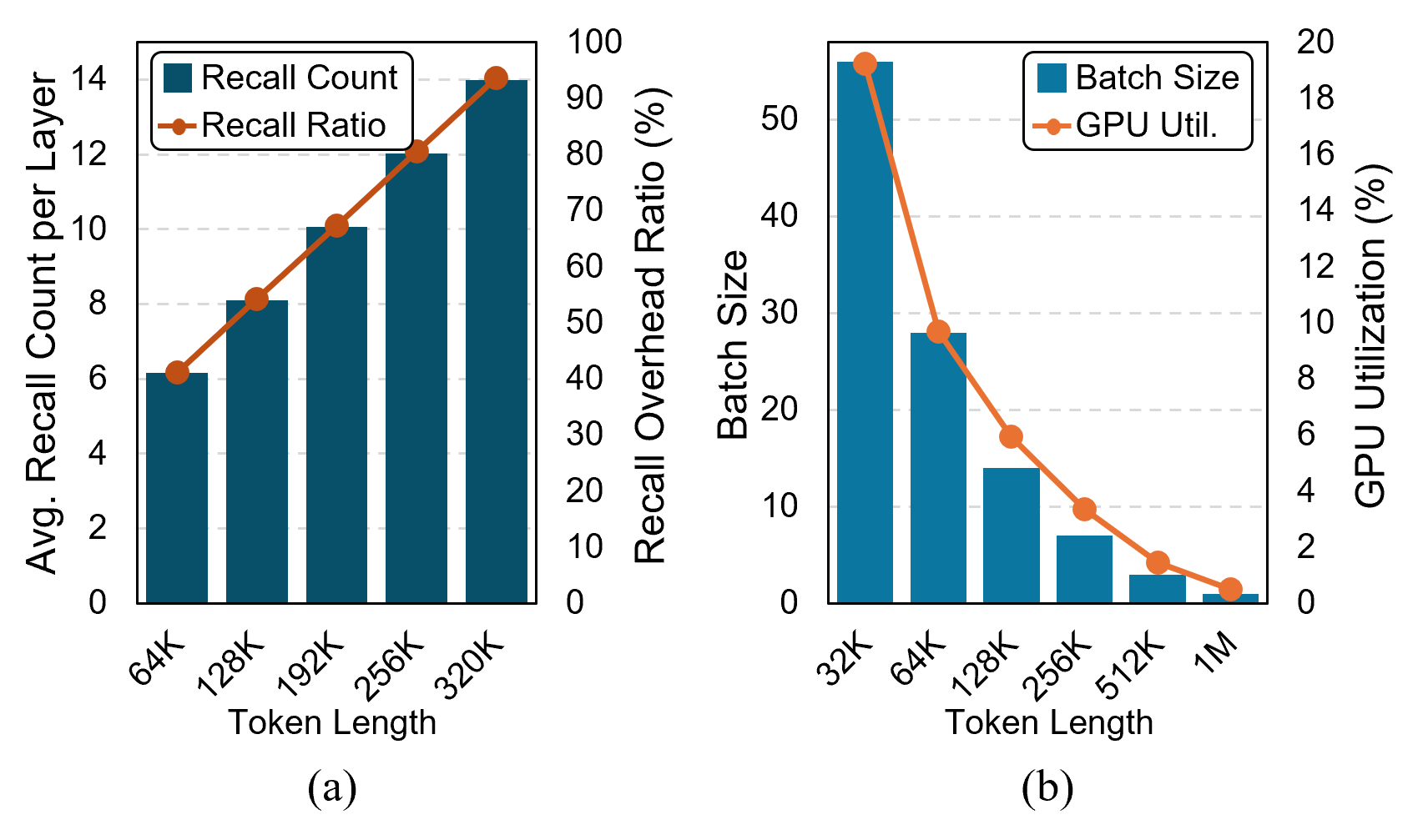}
    \vspace{-20pt}
    \caption{(a) As sequence length increases from 64 K to 320 K tokens, the average recall count per layer rises steadily, and the runtime recall overhead ratio compared to FC + attention latency also climbs, indicating that longer sequences incur significantly higher recall-induced overhead. (b) Poor FC computation efficiency on GPU under low batch size conditions. The left axis shows the maximum batch size and the right axis shows GPU utilization, both plotted against increasing sequence length (from 32K to 1M). Experiments were conducted on a single GPU (ArkVale) with Llama 3.1-8B. As sequence length grows, memory constraints reduce the achievable batch size, leading to a significant utilization drop
    }
    \label{fig:challenges}
\end{figure}

\textbf{\textit{Recall} Overhead in Long-Context Inference.}
As the context length grows, managing the KV-cache efficiently becomes more challenging not only in terms of capacity but also in retrieval overhead. In systems like ArkVale~\cite{chen2024arkvale}, where evicted KV-cache pages are backed up and selectively recalled from offloaded memory based on dynamic importance, longer context lengths naturally lead to a larger number of evicted pages. In other words, when estimating which pages of KV-cache are relevant for current queries, the likelihood of needing to recall additional pages from external memory increases. Our experimental analysis shown in Figure~\ref{fig:challenges}(a) reveals that the cost of recall scales with the total number of tokens and thus the number of managed pages. Although each recall operation incurs only a small transfer (approx. 5 MB) and low latency compared to overall decoding time, the cumulative recall activity becomes non-trivial for extremely long context exceeding hundreds of thousands of tokens. This growing recall overhead highlights the importance of designing scalable page summarization/compression methods and efficient importance estimation methods to sustain low-latency inference as context lengths continue to increase.

\textbf{GPU Underutilization.}
Note that in dynamic token selection algorithms, a sufficiently large number of tokens $T_{Budget}$ should be considered for GPU's attention computation to maintain the accuracy. Furthermore, $T_{Budget}$ grows as the context length increases, demanding more memory capacity. This growth in memory footprint, however, leads to a significant capacity bottleneck\textcolor{black}{, especially on GPU with limited memory constraints}: as the context length $T_{Budget}$ increases, the KV-cache size expands proportionally to $O(B \cdot T_{Budget}\cdot n_l\cdot n_h\cdot d_h)$. To stay within limited GPU's  memory, the system is forced to reduce batch size $B$, decreasing the number of concurrent requests. This constraint directly impacts throughput and latency in \textcolor{black}{FC computation of} long-context LLM inference. As shown in Figure~\ref{fig:challenges}(b), the longer context length constrains the batch size, thus the GPU utilization is degraded. These trends highlight the growing memory pressure imposed by longer context windows and underscore the need to reduce the GPU-side KV-cache to avoid batch size constraints and improve FC computation efficiency.

\section{Methodology}
\label{sec:methodology}

While CXL memory expansion extends the capacity of GPU systems for large-scale inference, it still incurs substantial data offloading overhead between the GPU and external CXL memory, particularly for long-context LLM inference that incurs a large number of recalls. Motivated by recent research exploring the augmentation of CXL memory expanders with processing-near-memory (PNM) capabilities~\cite{kwon2023failure,tangexploring,gouk2024breaking,park2024cxlpim,ham2024low,ji2024demystifying,gu2025pim}, we propose PNM-KV to enable lightweight computation for efficient KV-cache management directly near CXL memory, thereby supplementing GPU-based CXL memory expansion systems to avoid costly data offloading for scalable and efficient long-context LLM inference.

\begin{figure}[h]
    \centering
    \includegraphics[width=1\linewidth]{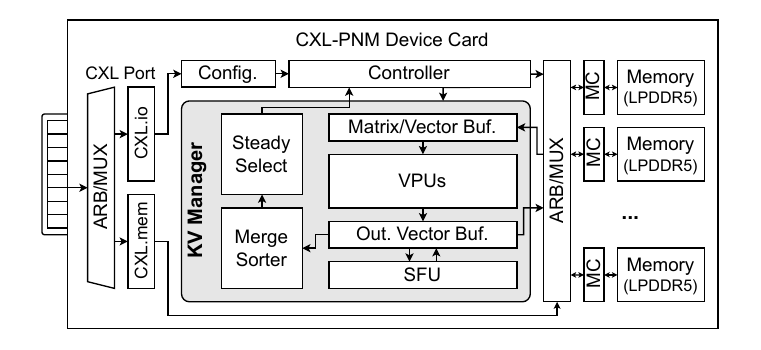}
    \vspace{-20pt}
    \caption{CXL-enabled PNM system overview.}
    \label{fig:lpddr-cxl-pnm}
\end{figure}

\subsection{PNM Architecture}
Figure~\ref{fig:lpddr-cxl-pnm} shows the overview of CXL-based PNM for efficient KV-cache management. The proposed CXL-PNM architecture realizes a scalable, high-capacity, and high-bandwidth memory expansion platform tailored for the LLM inference. Built upon an LPDDR5X-based CXL-PNM module~\cite{park2024cxlpim}, the system integrates dedicated components for KV-cache management between the CXL protocol stack and the LPDDR5X memory controllers. This design allows the device to be exposed as a standard CXL Type 3 memory device to the host CPU, while internally offering near-memory compute capabilities that can operate independently or in coordination with host-issued workloads.

\textbf{Control Mechanism.}
The CXL-PNM controller orchestrates memory access, instruction flow, and computation scheduling through two main interfaces and an internal pipeline. The CXL.mem interface allows both the host CPU and local accelerator to access LPDDR5X memory via AXI4-based arbitration, enabling efficient, low-latency sharing without contention common in DIMM-based PNMs. The CXL.io interface links the accelerator’s control unit to the host through an AHB interconnect for instruction programming and event signaling under PCIe standards. Internally, the control unit handles task scheduling and synchronization while maintaining metadata in a register file of ten 32-bit entries tracking key parameters such as layer count, sequence length, and memory mapping. A lightweight DMA engine provides high-bandwidth data transfer between LPDDR5X and 2.25 MB of internal vector/scalar buffers, reducing external traffic and improving throughput. Instruction execution follows a tightly coupled pipeline where instructions are fetched, decoded, scheduled, and dispatched to compute units, with a hardware scoreboard ensuring dependency tracking and efficient memory arbitration for scalable, autonomous LLM inference.

\begin{figure}[t]
    \centering
    \includegraphics[width=1\linewidth]{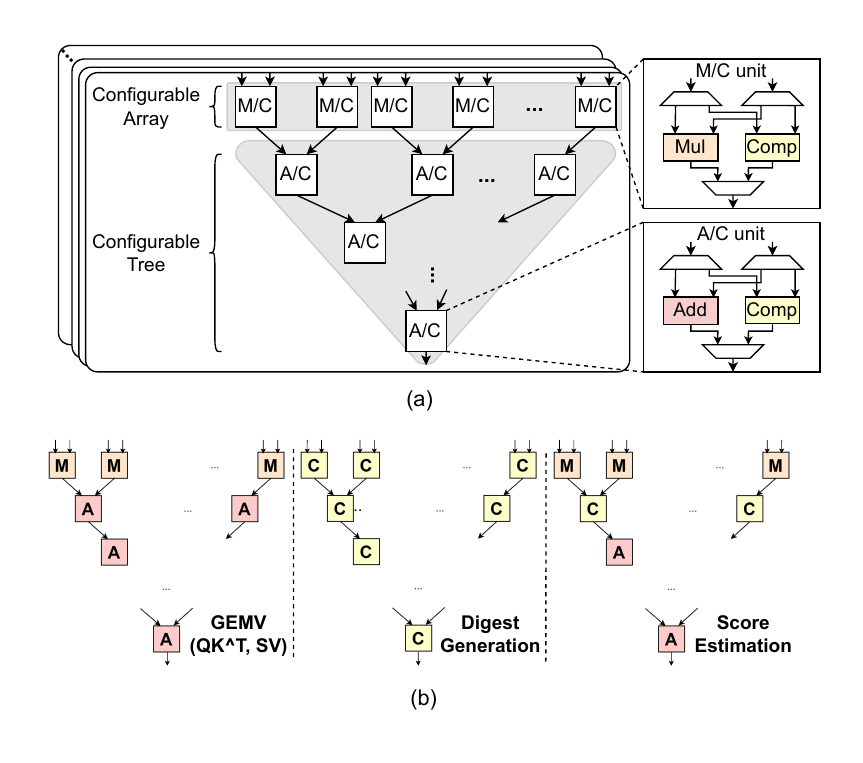}
    \vspace{-30pt} 
    \caption{Vector processing unit (VPU) for KV-cache management: (a) Hardware architecture of the VPU, (b) VPU configurations for three KV-cache management patterns.}
    \vspace{-5pt}
    \label{fig:vpu-arch}
\end{figure}

\textbf{KV-Cache Management Units.}
The computing subsystem of the proposed CXL-PNM architecture is optimized for efficient and scalable KV-cache management. The KV-cache management accelerates inference in long-context scenarios by dynamically selecting important token pages ($T_{Budget}$ tokens) for each query and performing attention only on the selected tokens. In this process, all tokens are partitioned into pages represented as compact digests, which are evaluated across queries through inner-product-based similarity and top-$K$ sorting to identify the most relevant tokens for attention computation. Unlike baseline PNM accelerators~\cite{park2024cxlpim} that integrate separate matrix and vector processing engines, our design focuses on decode-step kernels and employs a reconfigurable Vector Processing Unit (VPU) to support diverse computation patterns needed for KV-cache management with minimal area and energy overhead.

\begin{figure}[t]
    \centering
    \includegraphics[width=1\linewidth]{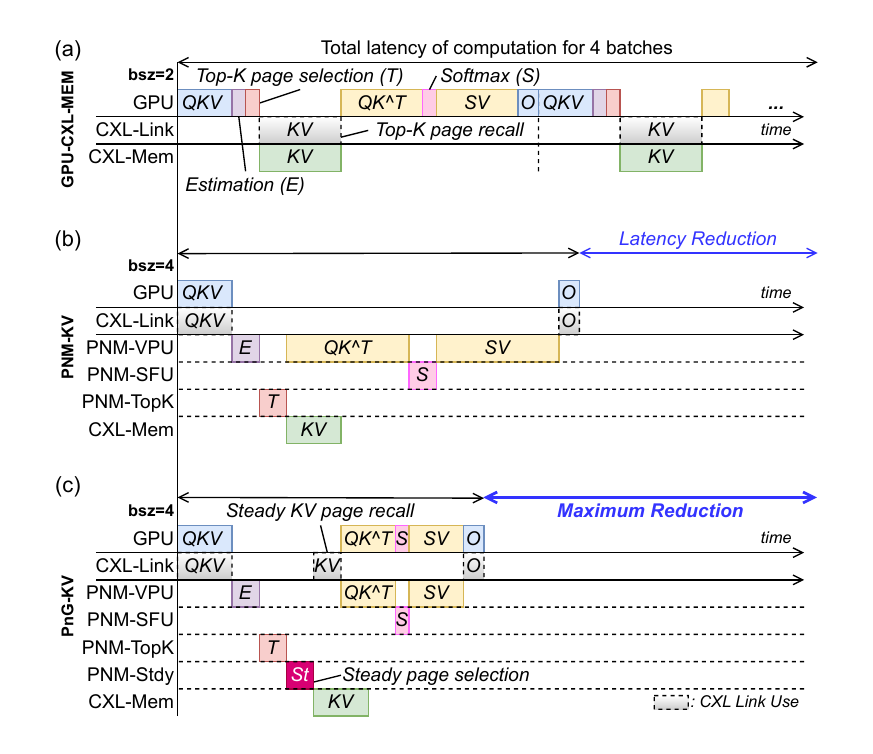}
    \vspace{-20pt}
    \caption{Baseline vs. CXL-PNM schemes. (a) GPU-CXL-Mem (Baseline): Tiny batch size wastes GPU compute and CXL memory. (b) PNM-KV: Move all batch data + attention to CXL-PNM—larger batches. (c) PnG-KV: Split attention between GPU and CXL-PNM to load-balance, fully leveraging CXL’s high-capacity memory and the GPU’s powerful compute resources.}
    \vspace{-5pt}
    \label{fig:GPU-PNM-timeline}
\end{figure}

Figure~\ref{fig:vpu-arch}(a) illustrates the microarchitecture of the Vector Processing Unit (VPU) for KV-cache management. The VPU is designed to operate flexibly in three primary modes: (1) General Matrix-Vector Multiplication (GEMV) mode for attention computations, (2) Digest Generation mode involving min/max comparisons for page summarization, and (3) Score Estimation mode based on vector-distance calculations. This configurability is enabled by dynamically programmable functional units within the VPU datapath, allowing efficient reuse of the same hardware resources 
across different stages of KV-cache management, as shown in Figure~\ref{fig:vpu-arch}(b). The VPU integrates standard vector ALUs alongside specialized comparison and distance computation units to accelerate these operations natively, eliminating the need for CPU offloading or redundant hardware duplication.

When examining each configuration in Figure~\ref{fig:vpu-arch}(b), the GEMV operations for the attention computations ($QK^T$ and $SV$) are mapped onto the VPU, similar to standard vector ALUs. The top configurable array is programmed as multipliers (M), while the underlying configurable tree is programmed as adders (A). Second, during digest generation, both the configurable array and tree are programmed as comparators (C) to extract the min/max values within each page, effectively summarizing the salient token features in that page. Finally, during the score estimation phase, the VPU identifies important pages for the current query by performing inner-product operations between the query and each page's digest. Since each digest contains two vectors representing the min and max features, the top array of the configurable tree is again programmed as comparators (C) to select the higher result between the inner products with the min and max vectors, thereby computing the final score.

\begin{figure*}[t]
    \centering
    \includegraphics[width=1\linewidth]{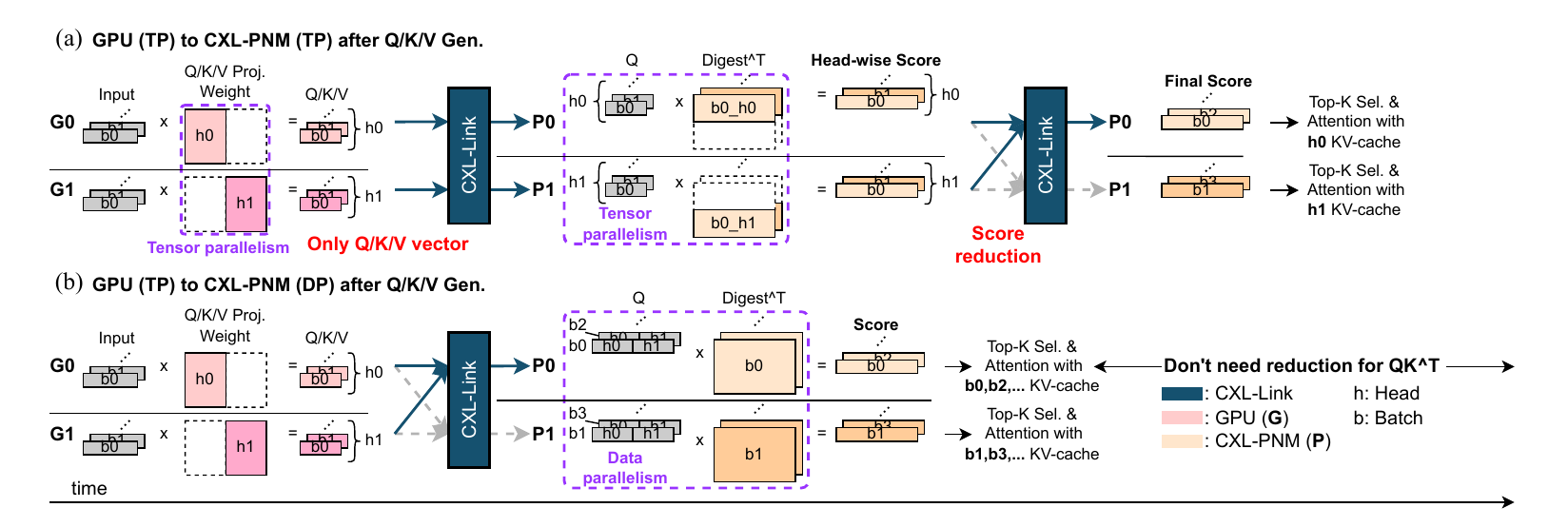}
    \vspace{-20pt}
    \caption{GPU to CXL-PNM parallel mappings: (a) TP→TP adds N $\times$ times Top-$K$ sorting and (N-1) $\times$ times reduction messages, while (b) TP→DP uses scatter-gather with the same data volume but no reductions. (N: \# of PNM Devices.)}
    \label{fig:TP-vs-DP}
\end{figure*}

In addition to the flexible VPU, the proposed CXL-PNM architecture includes a parallel Top-$K$ Sorter to expedite the retrieval of the most relevant token pages. The Top-$K$ Sorter implements a highly parallelized merge-sort algorithm, enabling fast extraction of Top-$K$ page scores across large token-page candidates. To execute softmax near memory, the Special Functional Unit (SFU) couples with the VPU. It integrates multiplier and adder arrays with vector length matched to the active VPU lanes, along with an adder tree for sum reduction and LUT-based exponent and reciprocal units.

Figure~\ref{fig:GPU-PNM-timeline} illustrates the execution schedules of the baseline (GPU with simple CXL memory, referred to as GPU-CXL-Mem) and the proposed PNM-KV and PnG-KV configurations for efficient KV-cache management. Compared to the GPU-CXL-Mem baseline (Figure~\ref{fig:GPU-PNM-timeline}(a)), the PNM-enabled KV-cache management effectively offloads the full KV-cache management and attention computation (Figure~\ref{fig:GPU-PNM-timeline}(b)). Operationally, the instruction pipeline fetches instructions from an instruction buffer, decodes them through a dedicated decoder, and schedules operations across the VPU and Top-$K$ Sorter using a hardware scheduler. A hardware scoreboard tracks inter-instruction dependencies, ensuring optimized instruction-level parallelism without pipeline stalls. Memory accesses for intermediate and final data are handled transparently through the CXL.mem interface, where arbitration logic dynamically balances host and accelerator memory requests to sustain high utilization.

\subsection{Workload Partitioning for Scalable PNM}
In the proposed CXL-PNM-powered GPU system, GPU focuses on FC computations while offloading the entire attention computation to CXL-PNM. This workload partitioning is designed to maximize scalability while minimizing communication overhead, a critical requirement for efficient multi-million-token inference. However, in multi-GPU and multi-PNM systems, additional overhead may arise depending on how the partitioned workloads are parallelized. To enable true scalability, we propose an optimal parallel mapping strategy between the GPU and CXL-PNM. Figure~\ref{fig:TP-vs-DP} illustrates the overall concepts of workload partitioning and parallelism strategy for multiple GPUs and PNMs.

In baseline GPU systems augmented with CXL memory expansion, tensor parallelism (TP) is commonly employed, where the workload is divided along the head dimension ($n_h$), which helps alleviate memory pressure by distributing the massive model parameters of LLMs across multiple devices (GPU0 (G0) → head0 (h0), GPU1 (G1) → head1 (h1)) as shown on the GPU-side in Figure~\ref{fig:TP-vs-DP}(a). Under TP, both FC and attention computations are deterministically partitioned across GPUs, and the associated KV-cache is also deterministically sliced and offloaded to CXL memory. This deterministic partitioning allows simple CXL memory expansion to support the workload without requiring dynamic coherence protocols, as each GPU accesses only its pre-assigned KV-cache partitions.

Building on this baseline, our system offloads attention computation to CXL-PNM while retaining FC computation on the GPU. This partitioning yields a key insight: during decoding, the only data exchanged between GPU and PNM is the batched activation output from the FC layers (Q, K, V vectors), which serves as the query for attention. The size of this activation depends solely on batch size and hidden dimension, independent of context length. Consequently, even as the context window expands to hundreds of thousands or millions of tokens, GPU–PNM data transfer volume remains constant. This context-length-independent communication prevents CXL link bottlenecks, enabling scalable inference for extremely long contexts.

While this partitioning enables scalable offloading, it raises new challenges for parallelizing attention computation across multiple PNM devices. In conventional GPU systems, TP is applied uniformly across both FC and attention computation to maximize GPU utilization. However, applying TP to attention computation on CXL-PNM devices would require a cross-device reduction that aggregates similarity scores across heads by concatenating bo\_h0 and bo\_h1 into the final b0 score before Top-$K$ token selection as shown in Figure~\ref{fig:TP-vs-DP}(a). This reduction would be highly inefficient in CXL-based infrastructures, which lack native device-to-device communication support and rely on host-mediated transfers.

To address this, we propose using data parallelism (DP) for attention computation within the PNM cluster. As shown in Figure~\ref{fig:TP-vs-DP}(b), under DP, each PNM independently processes different batched requests (PNM0 (P0) → batch0, batch2, ... (b0,..), PNM1 (P1) → batch1, batch3, ... (b1,..)), allowing each device to compute token similarities and perform Top-$K$ selection locally without requiring any inter-device communication. This eliminates costly reductions, simplifies synchronization, and aligns naturally with the context-independent activation-based communication pattern between GPUs and PNMs. By leveraging DP, the system achieves scalable and efficient attention computation across multiple PNMs without introducing new communication bottlenecks, ensuring the practicality of CXL-PNM-based KV-cache management for long-context LLM inference.

\newcommand{\baseline}[1]{\textcolor{gray}{#1}}

\begin{figure}[t]
    \centering
    \includegraphics[width=1\linewidth]{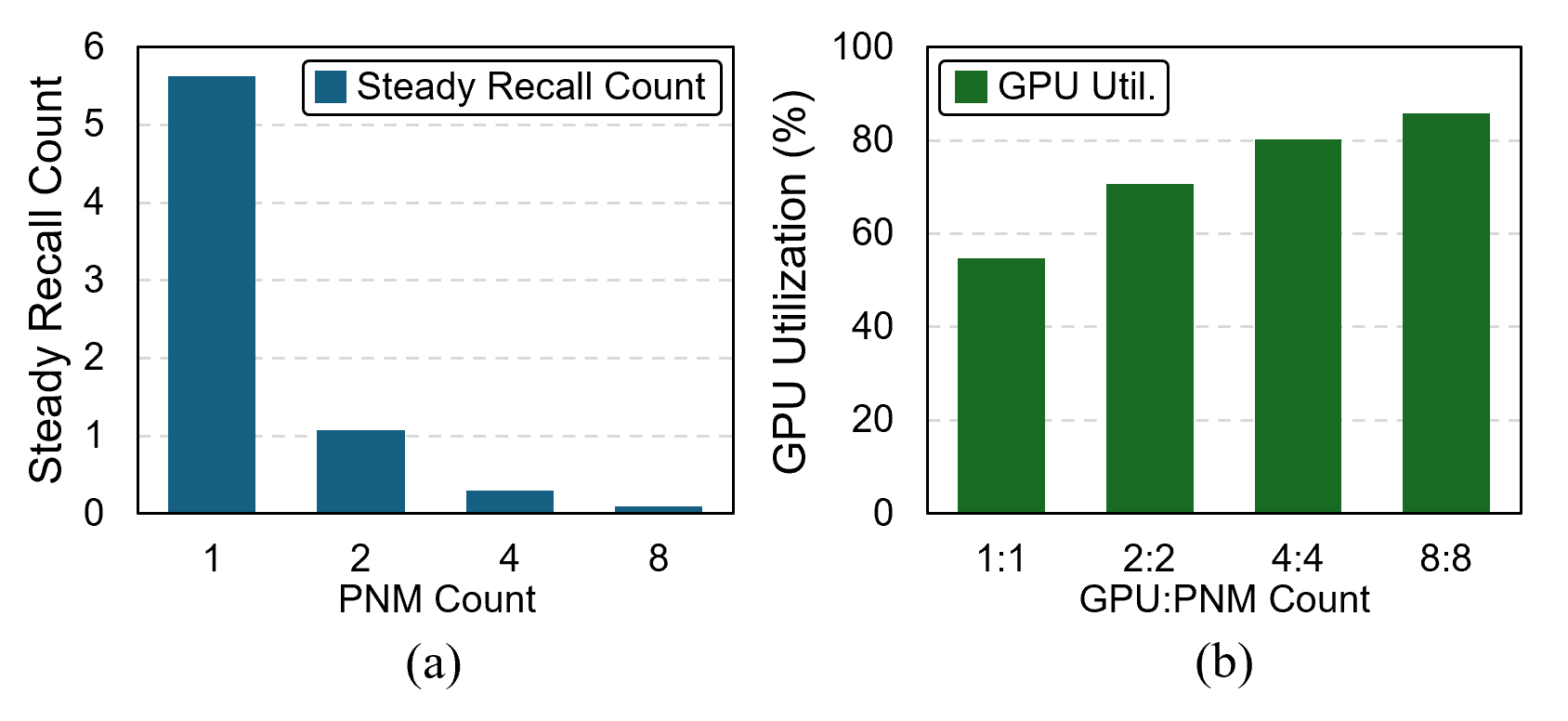}
    \vspace{-20pt} 
    \caption{Effects of Steady Selection on the \text{Llama3.1-8B} model. (a) As the number of PNM devices increases in single GPU, the number of recalls under Steady Selection drops sharply. (b) When both GPU and PNM counts are raised GPU utilization improves progressively.}
    \label{fig:steady-select-impact}
\end{figure}

\subsection{PNM-GPU Hybrid Execution}
\label{sec:steady}
While offloading the full attention computation and KV-cache management to CXL-PNM devices eliminates GPU memory bottlenecks and enables scalable long-context inference, it introduces a new inefficiency: GPU idleness during PNM attention computation. With the GPU only active during FC computation after receiving attention outputs, significant GPU resources remain underutilized, limiting overall system throughput. To address this issue, we propose PnG-KV, a hybrid attention computation model, where both the GPU and PNM collaboratively perform attention operations. In this model, a subset of KV-cache tokens is reassigned from the PNM to the GPU for local attention computation. Inspired by FlashAttention [11], we design attention to be parallelized independently across the GPU and PNM, thereby achieving accurate softmax computation by combining the exponential partial summations from both devices prior to the FC computation and updating the partial outputs based on the final summation. The partial attention outputs computed by the GPU and the PNM are then aggregated without requiring additional communication overhead, as the transfer of attention results is already part of the standard data flow between PNM and GPU for FC computation. Parallelizing with the power-hungry GPU increases instantaneous power consumption but shortens memory-bound decoding steps and raises end-to-end throughput, resulting in a clear throughput-energy trade-off.

However, enabling GPU-side attention computation reintroduces a critical challenge: \textit{token recall overhead}. Traditional GPU-based KV-cache management systems suffer from severe memory pressure because recalling a large number of tokens per sample ($T_{Budget}$) forces smaller effective batch sizes ($B_\mu$), degrading the efficiency of FC layers. A naïve reassignment would repeat this problem, reintroducing communication bottlenecks. To overcome this, we introduce \textit{Steady Token Selection}, a novel recall strategy that enables efficient GPU-side attention computation while preserving large batch execution. In Steady Token Selection, each GPU processes a full batch size $B$—the same as the PNM—thus maintaining FC layer throughput, but computes attention over only a small subset of recalled $T_{Steady}$ tokens per batch rather than the conventional $T_{Budget}$ tokens. 

The Steady Token Selection mechanism provides two key advantages. First, it minimizes memory transfer overhead, since recalling only a small $T_{Steady}$ subset significantly reduces the volume of data movement compared to full KV-cache recall. Second, it maintains the large batch size $B$ needed for efficient GPU execution, avoiding the micro-batching penalties of traditional GPU-based KV-cache systems. As shown in Figure~\ref{fig:steady-select-impact}(a), by leveraging the Steady Token Selection process described in Algorithm~\ref{alg:steady_token_selection}, increasing the number of PNMs leads to a larger effective batch size, while the proportion of steady tokens remains relatively small. This, in turn, decreases the recall overhead, making GPU participation lightweight and scalable even as the context length grows. Furthermore, Steady Token Selection becomes increasingly beneficial when scaling across multiple GPUs. As illustrated in Figure~\ref{fig:steady-select-impact}(b), assigning partial attention computation across more GPUs increases overall resource utilization and inference throughput, fully leveraging the available system parallelism while retaining the scalability advantages of PNM-based KV-cache management. The overall execution of the PNM-GPU hybrid execution (PnG-KV) is illustrated in Figure~\ref{fig:GPU-PNM-timeline}(c).

\begin{algorithm}[t]
\caption{KV-Cache Replacement ArkVale vs Steady Select}
\label{alg:steady_token_selection}
\begin{flushleft}
\textbf{Input:} $I$ – page importance scores\\ 
\textbf{Notation:} ${P}$ – on-GPU KV-cache\\
\hspace*{14mm}$r$ – recall KV-cache\\
\hspace*{14mm}$e$ – evict KV-cache\\
\end{flushleft}
\begin{algorithmic}[1]
    \State \textbf{1) Score Ranking:}
    \State \quad 
        ${S}\;\gets\;\textsc{Sort}_{\downarrow}(I)$
    \State \textbf{2) Page Select:}
    \baseline{
    \LineComment {ArkVale: \Comment{$T_{Budget} \times B_\mu\!\le\! |{P}|$}} 
    \State \hspace{1em} $r \gets S[:K] - P$ \Comment{recall: new Top-$K$ not in $P$}
    \State \hspace{1em} $e \gets (P - S[:K])[-|r|:]$ \Comment{evict: $|r|$ from lowest-score}
    }
    \LineComment {Steady-Select: \Comment{$T_{Steady \times B}\!\le\! |{P}|$}}
    \State \hspace{1em} $e \gets P - S[:T_{Budget}]$ \Comment{evict: not in budget set}
    \State \hspace{1em} $r \gets (S[:T_{Budget}] - P)[:|e|]$ \Comment{recall: |e| top pages not in P}
    \State \textbf{3) Evict \& Recall Pages:}
    \State \quad 
        ${P}[e]\gets\;r$ \Comment{overwrite evicted slots with recalled pages}
  \State \Return ${P}$
\end{algorithmic}
\end{algorithm}

Algorithm~\ref{alg:steady_token_selection} formalizes the proposed Steady Token Selection process, detailing how batch samples ($B$) and token subsets ($T_{Steady}$) are dynamically assigned during hybrid attention computation. Based on the GPU and PNM partitioning (PNM-KV), the on-GPU KV-cache ($P$) is allocated according to the remaining available GPU memory. Within $P$, $T_{Steady}$ tokens are stored through the Steady Token Selection process, where the number of tokens is determined to fit within the capacity that can accommodate all batch samples enabled by the proposed PNM-KV configuration. During the generation stage, at each step, from the $T_{Budget}$ tokens jointly processed by the GPU and PNM, the subset that remains steady is recalled to the GPU for attention computation. In this process, the GPU evicts page tokens that are no longer part of the current $T_{Budget}$ and replaces them with newly selected steady tokens. This hybrid execution model ensures that both GPU and PNM resources are effectively utilized without reintroducing memory bottlenecks, enabling scalable, high-throughput, multi-million-token inference across heterogeneous GPU–PNM architectures. 

\begin{figure}[h]
    \centering
    \includegraphics[width=1\linewidth]{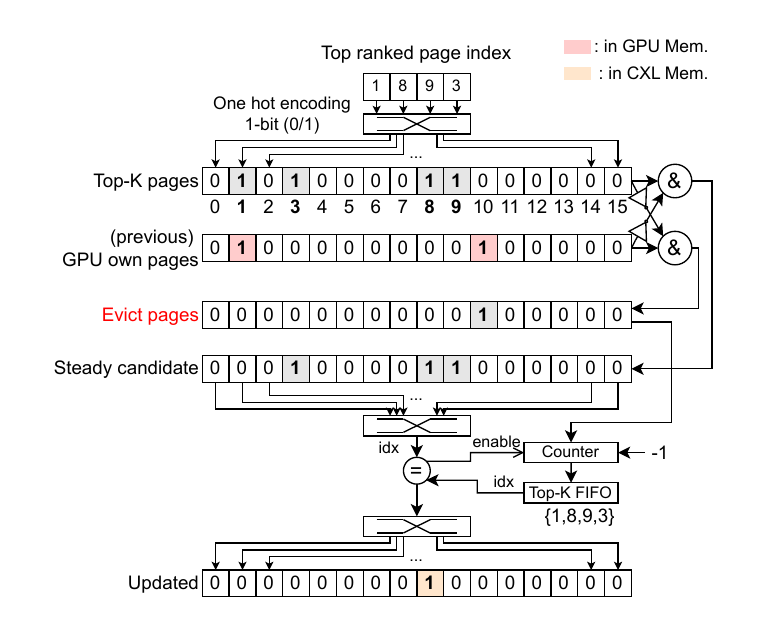}
    \vspace{-20pt} 
    \caption{Steady Selection toy example.}
    \vspace{10pt} 
    \label{fig:Steady-select}
\end{figure}

Figure~\ref{fig:Steady-select} further illustrates the micro-architecture of Steady selector, which includes bitmask-based page comparison to quickly process steady selection. The selector receives two masks: the Top-$K$ mask derived from the sorted list $S$ ($S[:T_{Budget}]$) and the current GPU-resident mask $P$. A bitwise AND between the complement of the Top-$K$ mask and the $P$ mask produces an eviction mask for entries not in the Top-$K$, whereas a bitwise AND between the complement of the $P$ mask and the Top-$K$ mask yields the recall candidates. A simple counter iterates the sorted Top-$K$ indices FIFO for exactly the number of eviction slots and compares with steady candidate page indices. It overwrites the enabled eviction slots in GPU memory with the corresponding recalled pages.
\section{Evaluation}
\label{sec:experiments}

\subsection{Evaluation Settings}
\label{subsec:experiments_settings}

\textbf{Software Stack.}
We develop a lightweight software stack to enable seamless integration of CXL-PNM into Python-based LLM inference programs, consisting of a CXL-PNM device driver and a Python runtime library. The driver extends the Linux DAX (Direct Access) framework to register CXL memory (CXL.mem) and accelerator control registers (CXL.io) as \texttt{/dev/dax0.0} and \texttt{/dev/mem}, respectively, allowing user-space programs to directly access memory regions using standard load/store instructions without explicit data transfers. The driver also exposes APIs for configuring device-specific registers and manages interrupts via the MSI-X (Message Signaled Interrupt eXtension) interface~\cite{msix_spec}. Upon task completion, the accelerator asserts an interrupt that is handled by an interrupt service routine (ISR). The Python runtime enables transparent offload without source changes by managing memory allocation within the CXL space, loading model parameters, configuring control registers, and orchestrating instruction execution behind familiar programming interfaces, thereby maximizing programmability, memory-sharing efficiency, and scalable deployment across GPU and PNM nodes.

\begin{table}[h]
\centering
\caption{CXL-PNM platform architecture and operating parameters. The power consumption of the CXL-PNM controller comprises that of the CXL IPs and the LLM accelerator.}
\label{tab:cxl-pnm-arch}
\small
\begin{tabularx}{\linewidth}{l|X}
\hline
\textbf{\# of adder-tree multipliers/adders} & 4,096/4,064 (peak 8 TFLOPS) \\
\textbf{\# of comparators} & 8,160 \\
\textbf{On-chip buffer} & 2.25 MB \\
\textbf{DMA Buffers} & 1 MB \\
\textbf{I/O width of DRAM/SRAM} & 1,024/16,384 \\
\hline
\textbf{Technology/Frequency/Voltage} & 7 nm/1.0 GHz/1.0V \\
\textbf{CXL-PNM controller max power} & $\sim$90 W \\
\textbf{DRAM total power} & $\sim$40 W \\
\textbf{CXL-PNM platform total power} & $\sim$150 W \\
\hline
\end{tabularx}
\end{table}

\begin{figure*}[th!]
    \centering
    \includegraphics[width=1\textwidth]{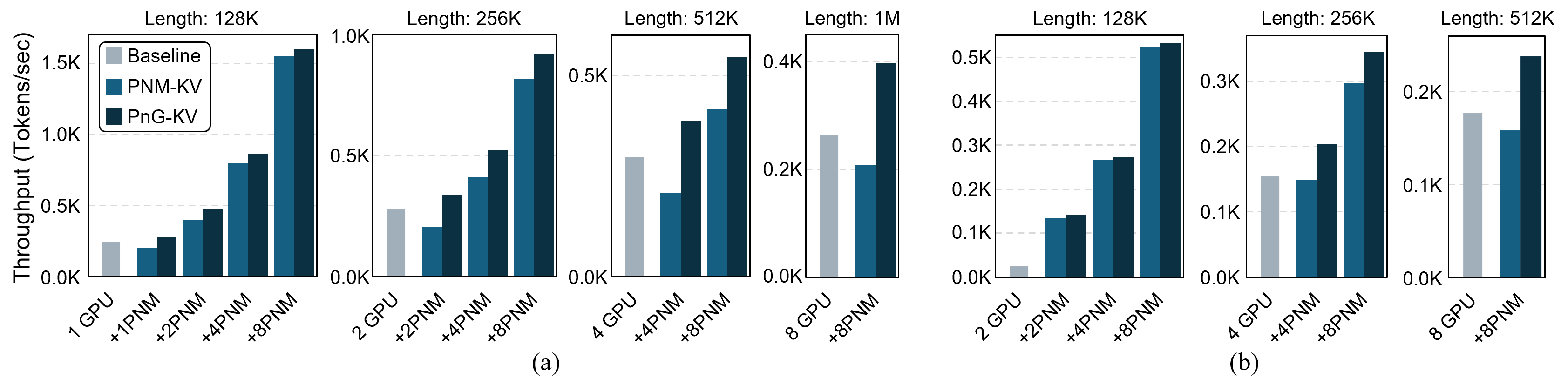}
    \vspace{-25pt}
    \caption{Throughput comparison with In-server PNM with KV-cache offloading. (a) Llama3.1-8B with context length from 128K (1GPU) to 1M tokens (8GPUs). (b) Llama3.1-70B with context length from 128K (2GPUs) to 512K tokens (8GPUs).}
    \label{fig:main1}
\end{figure*}

\begin{figure*}[th!]
    \centering
    \includegraphics[width=1\textwidth]{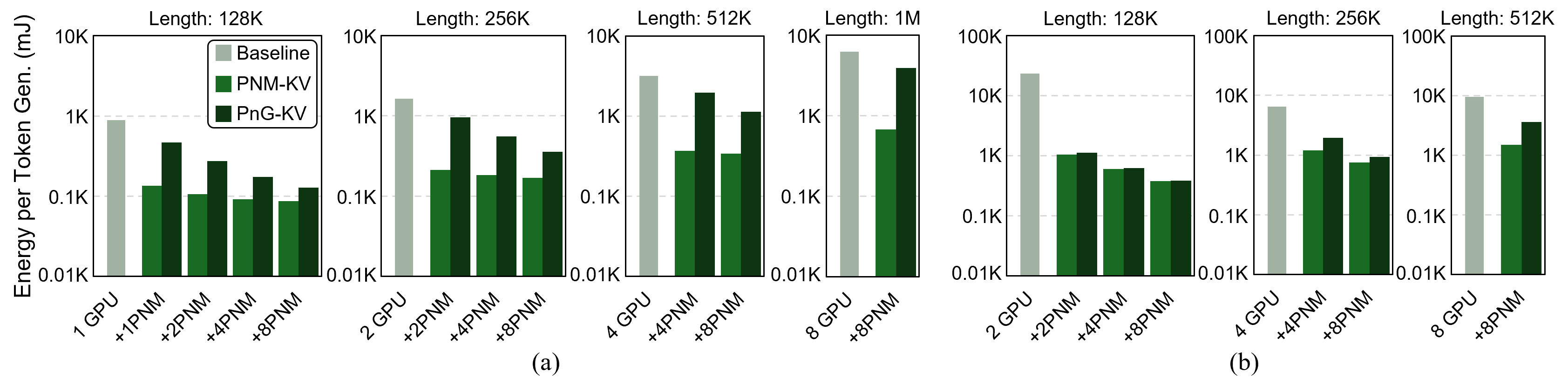}
    \vspace{-25pt}
    \caption{Energy consumption per token generation with In-server PNM with KV-cache offloading. (a) Llama3.1-8B with context length from 128K (1GPU) to 1M tokens (8GPUs). (b) Llama3.1-70B with context length from 128K (2GPUs) to 512K tokens (8GPUs).}
    \label{fig:main2}
\end{figure*}

\textbf{Hardware Implementation.}
We implement CXL-PNM as a custom CXL module interfacing LPDDR5X DRAM and integrate it into a server based on an Intel 4th-generation Xeon CPU. Architectural and operational details are summarized in Table~\ref{tab:cxl-pnm-arch}. In the CXL-PNM controller, the near-memory LLM accelerator consists of 32 VPU tiles. Each VPU instance integrates a 128-wide multiplier array, a 128-wide comparator array, a 64-length adder tree, and a 64-length comparator tree. All compute units operate in FP16 precision. The entire system is implemented as a 7\,nm ASIC operating at 1.0\,GHz. The typical power envelope of a CXL-PNM device is approximately 150\,W, with about 90\,W attributed to the controller and accelerator logic~\cite{dally2020gtcchina} and about 40\,W to LPDDR5X DRAM modules~\cite{chang2023sarc101}. For performance evaluation, we use a cycle-level simulator parameterized by synthesized timing and power models of the 7\,nm design. This design enables high compute and memory efficiency, reducing \textit{total cost of ownership} relative to traditional multi-GPU solutions.

\textbf{Evaluation Methodology.}
We report decode-stage performance of the CXL-PNM platform (1.1~TB/s LPDDR5X bandwidth, 512GB per module) against an NVIDIA DGX appliance populated with up to eight A100-80GB GPUs (2~TB/s HBM2E bandwidth)~\cite{A100}. We use Llama models~\cite{llama_models}: Llama3.1-8B fits within a single GPU’s memory, Llama3.1-70B requires at least two GPUs, and Llama3.1-405B demands two fully populated DGX systems. Inference uses NVIDIA FasterTransformer~\cite{fastertransformer}, which supports tensor and pipeline parallelism across multiple GPUs. Workloads span extreme long-context inference with 128K–1M tokens, representative of datacenter text-generation services~\cite{brown2020language}, on InfiniteBench En.Sum (longbook summarization)~\cite{zhang-etal-2024-bench}. For PnG-KV, the steady-token ratio is set empirically to $\#\mathrm{GPU}$/$(\#\mathrm{GPU}{+}\#\mathrm{PNM})$ to balance GPU-PNM work; higher ratios help until recall overhead dominates and throughput declines.

\subsection{Results}
\label{subsec:experiments_results}
\textbf{Server-Level Evaluation.}
Figure~\ref{fig:main1} compares the throughput across three In-server configurations: (i) a CXL–memory–expanded DGX baseline, (ii) a CXL-PNM–attached DGX system that offloads KV-cache from GPUs (PNM-KV), and (iii) the same PNM-KV with the hybrid steady mode enabled (PnG-KV) evaluated on (a) Llama3.1-8B and (b) Llama3.1-70B. Each bar group corresponds to a fixed number of GPU cards, varying the number of PNM devices. From left to right, both the number of GPU cards and context length (starting from 128K tokens) increase. Across all setups, increasing the number of PNM devices significantly boosts throughput, highlighting the importance of efficient KV-cache management for large-batch inference. The baseline quickly saturates due to GPU memory limits, while PNM-KV alleviates this by offloading KV-cache operations. PnG-KV further improves throughput by utilizing idle GPU compute during attention, particularly when PNM resources are limited. The unusually low 2-GPU baseline for Llama3.1-70B stems from a tiny batch imposed by model size and KV capacity at 128K context. When the number of PNM devices matches the number of GPUs, throughput can fall below the baseline because the batch-size gain in the FC layers is smaller than the attention throughput loss caused by the lower bandwidth of CXL–PNM relative to the GPU. 

As shown in Figure~\ref{fig:main2}, the trade-off of PnG-KV becomes clear. It delivers higher throughput than PNM-KV by on average 28.3\%, while its energy can be higher by about 2.2$\times$ since both GPU and PNM are active. 
Compared with the baseline, PnG-KV still achieves 2.7$\times$ higher throughput and 4.9$\times$ lower energy on average. 
For Llama3.1-70B inference, a 2GPU+8PNM PnG-KV configuration achieves up to 60$\times$ lower energy per token than the 2GPU baseline. This is due to two factors. First, increasing the batch size from 1 to 96 lifts GPU FC-layer efficiency by as much as 96$\times$. Second, offloading attention to 8$\times$ PNM devices yields 20$\times$ better efficiency due to reduced data movement and low-power LPDDR5X.

\begin{figure}[t]
    \centering
    \includegraphics[width=1\linewidth]{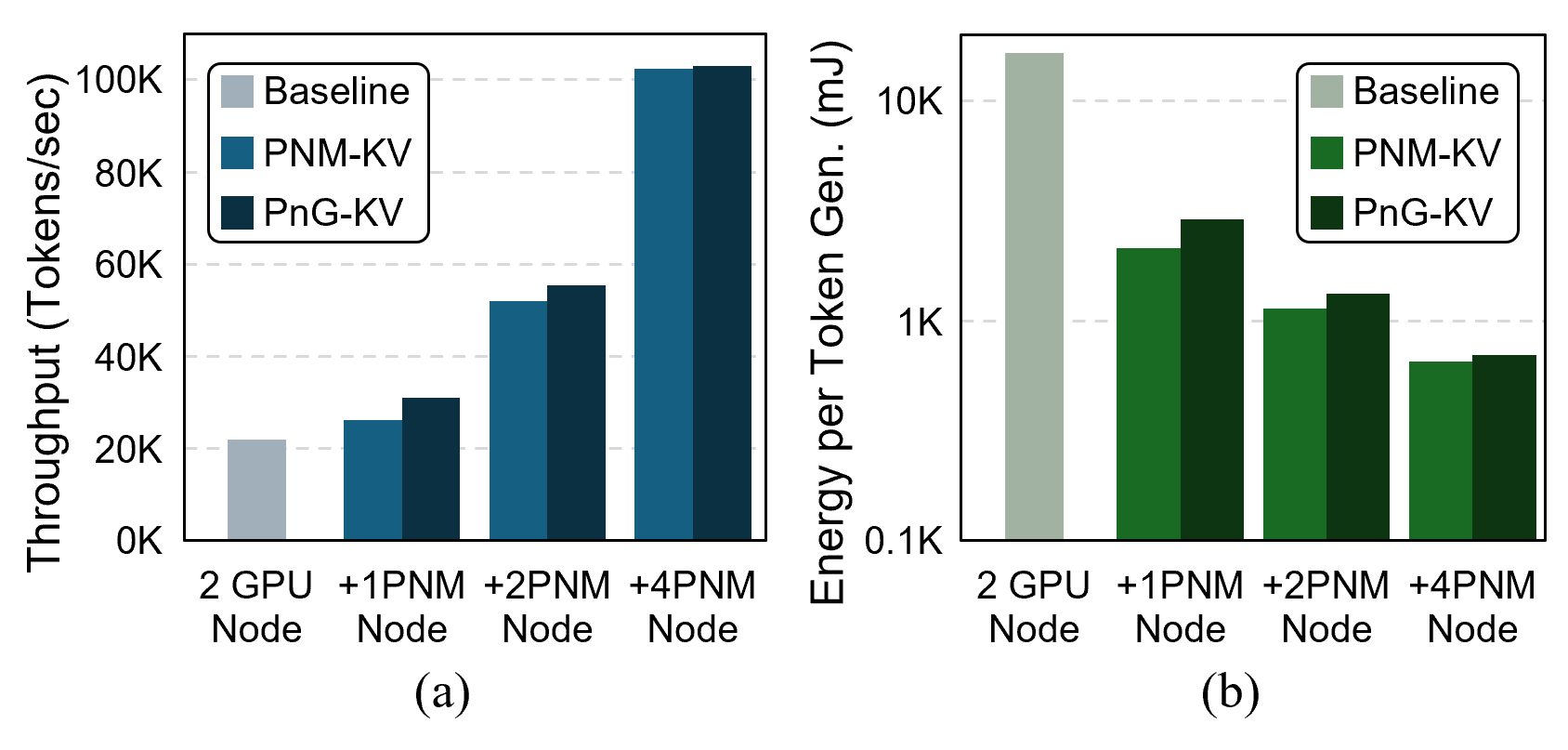}
    \vspace{-25pt}
    \caption{Rack-scale PNM with KV-cache offloading with 1M tokens. (a) Throughput comparison. (b) Energy consumption comparison.}
    \vspace{-15pt}
    \label{fig:throu_energy}
\end{figure}

\begin{figure*}[t]
    \centering
    \includegraphics[width=1\textwidth]{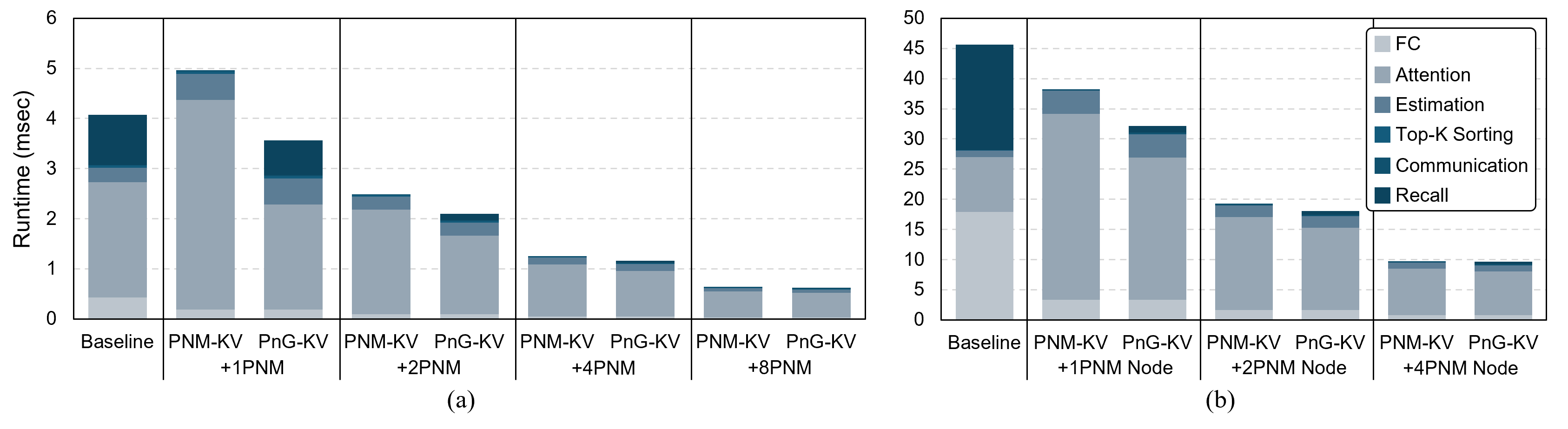}
    \vspace{-20pt}
    \caption{Breakdown of per-token latency. 
(a) Evaluated on Llama3.1-8B at 128K context length with a single GPU card. (b) Evaluated on Llama3.1-405B at 1M context length with a 2$\times$GPU node.}
    \label{fig:main5}
\end{figure*}

\textbf{Rack-Scale Evaluation.}
Figure~\ref{fig:throu_energy}(a)
shows the throughput results for rack-scale configurations, where a GPU node denotes a baseline DGX system consisting of 8$\times$A100-80GB GPUs, and a PNM node denotes a server with 16$\times$CXL-PNM devices (PNM-KV). We evaluate on Llama3.1-405B with a 1M-token context using 2$\times$GPU nodes while scaling PNM nodes from 1 to 4. In memory-capacity constrained settings with long context where even two GPU nodes are insufficient, batch sizes collapse and GPU utilization drops. Adding PNM nodes augments capacity, keeping more KV resident, so attention can be parallelized and throughput scales while KV-cache management becomes more efficient.

Figure~\ref{fig:throu_energy}(b) 
compares energy consumption, where both PNM-KV and PnG-KV outperform the baseline and the same throughput-energy trade-off of PnG-KV appears at rack-scale. 
As with server-level experiments, PnG-KV shows early gains about 8.5\% higher throughput than PNM-KV, while consuming about 19.3\% more energy. 
It remains more efficient than the baseline (2.6$\times$ higher throughput and 11.8$\times$ lower energy) even as fully distributed near-memory processing becomes dominant as the PNM cluster grows.

\begin{table}[b]
\centering
    \vspace{-5pt}
    \caption{Comparison of operating costs and hardware cost. Electricity rates are based on the March 2025 Consumer Price Index, with a unit price of 0.181\$/kWh~\cite{gu2025pim}.}
    \vspace{-5pt}
    \resizebox{\linewidth}{!}{
        \begin{tabular}{c|c|c}
            \toprule
            \cline{1-3}
            
            \textbf{Hardware}          & \textbf{A100 GPU}    & \textbf{CXL-PNM} \\
            \hline
            Max power per device      & $\sim400~\mathrm{W}$     & $\sim150~\mathrm{W}$ \\
            Operating cost per device          & 0.072\$/hour   & 0.027\$/hour \\
            Hardware cost per device       & 0.761\$/hour & 0.266\$/hour \\
            \hline
            \bottomrule
        \end{tabular}
    }
    \vspace{-10pt}
\label{tab:hardware_cost}
\end{table}

\textbf{Runtime Breakdown.}
Figure~\ref{fig:main5} presents the runtime breakdown of per-token latency for Llama3.1-8B (a) and Llama3.1-405B (b) at 128K context length, using a single GPU card while scaling PNM cards from 1 to 8 and 1 to 4, respectively. In the single GPU baseline, attention computation and KV-cache recall dominate latency. With PNM-KV, all attention runs in PNM, eliminating recall and enabling larger batches that improve FC efficiency. However, attention itself can be slower on PNM due to its lower memory bandwidth than the GPU. With PnG-KV, steady tokens are served on the GPU in parallel, which reintroduces some recall but reduces end-to-end latency by utilizing otherwise idle GPU resources. As we add more PNM devices, the feasible batch size increases and more steady tokens remain resident on the GPU, shrinking recall overhead and further lowering total latency. In the 2$\times$GPU node baseline at a 1M-token context, the absolute number of recalled tokens is much larger and data-transfer latency becomes dominant. Scaling out PNM nodes increases near-memory bandwidth and batches together, accelerating attention and FC near-linearly and reducing recall overhead, overall runtime falls markedly compared with traditional CXL-memory-expanded GPU system offloading approaches. In both cases, hidden-state communication and Top-$K$ sorting contribute negligible latency across all configurations compared with attention, recall, and FC.

\begin{figure}[th!]
    \centering
     \includegraphics[width=1\linewidth]{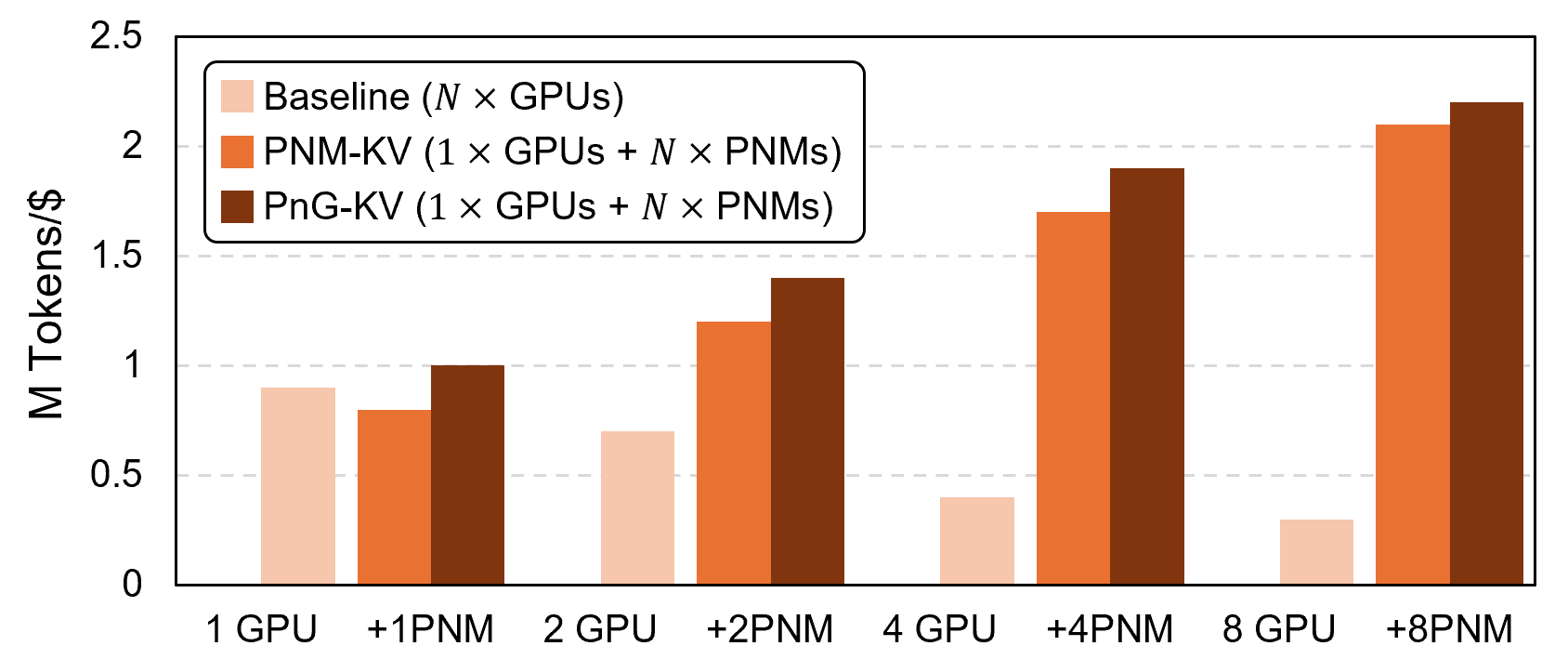}
    \vspace{-15pt}
    \caption{Comparison of total cost efficiency between GPU scaling and CXL-PNM scaling for Llama3.1-8B, 128K.}
    \label{fig:main6}
\end{figure}

\textbf{Total Cost of Ownership.}
Figure~\ref{fig:main6} compares total cost efficiency using throughput-per-dollar for Llama3.1-8B under a 128K context. The operating and hardware costs are based on Table~\ref{tab:hardware_cost}. We use GPU scaling as the baseline and contrast it with CXL-PNM scaling in a single-GPU scenario. Despite the high cost of GPUs, naive GPU scaling encounters memory-capacity limits and rising recall overhead, causing throughput saturation and reducing total cost efficiency (e.g., to ~0.4$\times$ relative to a single GPU). In contrast, holding a single GPU and scaling PNM devices increases total cost efficiency monotonically, primarily because PNM eliminates recall and, with large memory capacity and sufficient bandwidth, delivers nearly linear throughput scaling, while also offering lower cost and lower energy consumption. With the hybrid PnG-KV strategy, 1GPU+8PNM achieves up to 7.3$\times$ higher cost efficiency than the 8-GPU baseline. Although PnG-KV is less energy-efficient than PNM-KV due to power-intensive GPU participation in attention, its throughput gain with low recall overhead yields the best tokens-per-dollar. While GPUs remain optimal for dense FC computation, PNMs provide superior total cost efficiency for sparse attention under long contexts, and the hybrid GPU–PNM model balances the compute and memory-intensive phases of LLM inference.

\section{Related Work}
\label{relatedwork}
\subsection{KV-Cache Management}
\textbf{KV Cache Selection Methods.}
Efficient KV-cache management is critical for scaling LLM inference to long contexts. Recent advances fall into two families: methods that permanently evict unneeded tokens~\cite{xiao2023efficient,han2023lm,zhang2023h2o,liu2023scissorhands,chen2024nacl,adnan2024keyformer,kim2024infinipot} and methods that dynamically select tokens without discarding them~\cite{chen2024arkvale,xiao2024infllm,tang2024quest,zhang2024pqcache,hooper2024squeezed,liu2024retrievalattention,fountas2024human,liu2024clusterkv,ribar2023sparq}. Eviction-based approaches reduce memory usage by discarding less-important tokens, with examples like StreamingLLM using a sliding window and H2O employing a hierarchical cache. InfiniPot continuously distills essential context into a fixed-size cache. Their main drawback is irreversibility—evicted tokens cannot be recalled, possibly degrading accuracy. In contrast, non-eviction methods such as ArkVale, InfLLM, Quest, and SqueezedAttention preserve all tokens in secondary memory while dynamically selecting relevant ones, yielding higher accuracy at the expense of increased capacity and retrieval overhead.

\textbf{Dynamic Token Retrieval Strategies.}
In non-eviction frameworks, efficient token retrieval is key. Many methods organize the KV-cache into chunks or clusters with compact summaries retained on the GPU. At inference, these summaries guide query-to-summary matching to select the most relevant pages. InfLLM uses pooled key representations; Quest computes fast relevance scores based on key statistics; SqueezedAttention pre-clusters the context and selects clusters via centroid similarity; ArkVale adaptively chooses pages per query; and ClusterKV groups semantically similar tokens into recallable clusters. These strategies preserve full-context coverage while enabling scalable inference over millions of tokens. Our work presents an efficient KV-cache system that minimizes retrieval latency at extreme scales.

\textbf{KV-Cache Management System.}
Recent system-level research focuses on enhancing memory efficiency through sparsity exploitation, offloading, and quantization. ALISA~\cite{zhao2024alisa} accelerates inference by caching only salient tokens via sparsity-aware attention and dynamic scheduling. InfiniGen~\cite{lee2024infinigen} reduces offloading overhead through speculative prefetching, while NEO~\cite{jiang2024neo} employs an asymmetric GPU–CPU pipeline to support memory-bound workloads. Oaken~\cite{kim2025oaken} uses hybrid offline–online KV quantization to boost capacity. While these methods leverage CPU memory as an extension of GPU memory, they remain limited by fixed CPU-attached memory bandwidth and capacity. In contrast, our work utilizes CXL-enabled memory expansion to scale capacity and enable near-memory processing, overcoming bandwidth bottlenecks and supporting efficient, scalable KV-cache management.

\subsection{CXL-enabled Memory System}
\textbf{CXL System Characterization and Realization.}
Compute Express Link (CXL) is emerging as a key enabler for disaggregated and tiered memory systems, providing coherent, low-latency access to external memory devices. Extensive research has characterized CXL-based architectures~\cite{li2023pond,gouk2023memory,yang2022performance,zeng2025performance,sun2023demystifying,mao2024cxl,wang2024exploring,yang2025architectural,jang2024bridging,berger2025octopus}. For example, Pond~\cite{li2023pond} presents a production-ready pooling system that reduces cloud DRAM costs by up to 7\% with minimal performance loss, while DirectCXL~\cite{mao2024cxl} eliminates RDMA overhead using a low-latency, software-minimal pooling approach. Sun et al.~\cite{sun2023demystifying} compare genuine CXL systems to NUMA-based emulation and introduce a CXL-aware memory allocator, and Yang et al.~\cite{yang2025architectural} propose MIKU to mitigate DRAM interference in tiered memory. These studies provide a strong foundation for deploying CXL in modern memory systems.

\textbf{CXL for Near-Data Processing and AI Acceleration.}
Beyond memory pooling, CXL’s composability supports near-data processing (NDP) and scalable AI workloads. Recent efforts attach programmable compute logic or AI accelerators to CXL memory expanders, reducing data movement and boosting efficiency~\cite{kwon2023failure,tangexploring,gouk2024breaking,park2024cxlpim,ham2024low,ji2024demystifying,gu2025pim}. M²NDP~\cite{ham2024low} proposes a memory-mapped NDP architecture with lightweight host-device interaction and microthread execution for flexible offloading. Ji et al.~\cite{ji2024demystifying} analyze cooperative host-device execution in CXL Type-2 devices, highlighting coordination bottlenecks. Tang et al.~\cite{tangexploring} demonstrate that offloading LLM KV caches to CXL memory can maintain latency while reducing GPU usage by up to 87\%, and Gouk et al.~\cite{gouk2024breaking} develop a sub-100 ns latency CXL memory controller for seamless GPU expansion. Notably, CXL-PNM~\cite{park2024cxlpim} and CENT~\cite{gu2025pim} introduce CXL-enabled GPU-free systems for LLM inference, enabling large transformer models at lower cost and power. These advances position CXL as both a memory extension interface and a compute-capable backbone for next-generation AI systems.

\section{Conclusion} 
This paper presents a scalable CXL-enabled Processing-Near-Memory (PNM) architecture to address the growing memory and compute challenges in 1M-token long-context LLM inference. By offloading KV-cache management and attention computation to PNM devices directly attached via CXL, our design eliminates costly recall overheads, alleviates GPU memory pressure, and enables significantly larger batch sizes for high-throughput execution. We further propose a hybrid execution model combining GPU and PNM processing through steady-token selection, maintaining large batch FC computation while minimizing communication overhead. Implemented atop a custom CXL-PNM platform, our system demonstrates consistent performance and energy efficiency gains across server- and rack-scale deployments for models up to 405B parameters. Cost-wise, scaling PNMs yields higher cost efficiency, and the PnG-KV further improves tokens-per-dollar by exploiting otherwise idle GPU resources to raise throughput, despite higher operating power. These results highlight the potential of CXL-enabled multi-PNM systems as a scalable and efficient backbone for future long-context LLM inference workloads.

\begin{acks}
\textcolor{black}{This work was supported by the National Research Foundation of Korea (NRF) grant funded by the Korea government (MSIT) (Nos. RS-2025-00561961, RS-2023-00260527 and RS-2020-NR047143); the Institute of Information \& communications Technology Planning \& Evaluation (IITP) grant funded by the Korea government (MSIT) under the Artificial Intelligence Semiconductor Support Program to Nurture the Best Talents (IITP-2025-RS-2023-00253914) and No. 2022-0-00971 (Logic Synthesis for NVM-based PIM Computing Architecture); and Samsung Electronics Co., Ltd. (No. IO201210-08015-01), and the authors thank the Samsung Memory Research Center (SMRC) for providing research facilities and infrastructure.}
\end{acks}

\bibliographystyle{ACM-Reference-Format}

\bibliography{references}

\end{document}